\documentclass[twocolumn]{article} 
\usepackage{graphicx} 

\long\def\comment#1{}

\makeatletter
\def\@normalsize{\@setsize\normalsize{10pt}\xpt\@xpt
\abovedisplayskip 10pt plus2pt minus5pt\belowdisplayskip
\abovedisplayskip \abovedisplayshortskip \z@
plus3pt\belowdisplayshortskip 6pt plus3pt
minus3pt\let\@listi\@listI}

\def\subsize{\@setsize\subsize{12pt}\xipt\@xipt}
\def\section{\@startsection {section}{1}{\z@}{1.0ex plus
1ex minus .2ex}{.2ex plus .2ex}{\large\bf}}
\def\subsection{\@startsection
   {subsection}{2}{\z@}{.2ex plus 1ex} {.2ex plus .2ex}{\subsize\bf}}
\makeatother

\begin{document}

\date{}

\title{\huge \bf {Fast GPU Implementation of Sparse Signal Recovery from Random Projections}}

\author{M. Andrecut
\thanks{Manuscript received November 24, 2008. 
Institute for Biocomplexity and Informatics, 
University of Calgary, 
2500 University Drive NW, Calgary, 
Alberta, T2N 1N4, Canada, Email: mandrecu@ucalgary.ca. 
 }
}

\maketitle
\thispagestyle{empty}


{\hspace{1pc} {\it{\small Abstract}}{\bf{\small---We consider the problem 
of sparse signal recovery from a small number of
random projections (measurements). This is a well known NP-hard to solve
combinatorial optimization problem. A frequently used approach is based on
greedy iterative procedures, such as the Matching Pursuit (MP) algorithm.
Here, we discuss a fast GPU implementation of the MP algorithm, based on the
recently released NVIDIA CUDA API and CUBLAS library. The results show that
the GPU version is substantially faster (up to 31 times) than the highly
optimized CPU version based on CBLAS (GNU Scientific Library).

}}
 }


\section{Introduction}
\label{Introduction}

Recently there has been an increasing interest in recovering sparse signals
from their projection onto a small number of random vectors (see \cite{1}-\cite{7} and
the references within). Sparse signal expansions represent or approximate a
signal using only a small number of elements from a given basis or
dictionary. Unfortunately, the sparse recovery problem is known to be a
NP-hard combinatorial optimization problem, requiring the enumeration of all
possible collections of elements in a dictionary and searching for the
smallest collection which best approximates the signal. Several sub-optimal
methods have been recently developed \cite{1}-\cite{7}, such that a wide range of
applications have benefited from the progress made in this area. These
methods show good performance in finding the solution of the sparse
approximation problem. However, their major shortcoming resides in achieving
sufficient computational speed, which limits their application to difficult
real world applications which require heavy computational load. The recent
improvements in performance of graphics hardware have made Graphics
Processing Units (GPUs) strong candidates for approaching this problem.
Recently, NVIDIA has released a general-purpose oriented API for its
graphics hardware, called CUDA \cite{8}. In addition, NVIDIA has developed CUBLAS
which is a GPU optimized version of BLAS library (Basic Linear Algebra
Subroutines) built on top of CUDA \cite{9}. In this paper we discuss and evaluate
the CUBLAS implementation of the Matching Pursuit (MP) algorithm \cite{10}.
Although MP is not the most accurate algorithm for sparse signal recovery,
it is still the fastest and most frequently used in practice, due to its
relative simple formulation. Our numerical results show that the GPU version
of MP is substantially faster (up to 31 times) than the highly optimized CPU
version based on CBLAS (GNU Scientific Library) \cite{11}.

\section{The Sparse Recovery Problem}
\label{The Sparse Recovery Problem}

A high dimensional vector (signal) $\mathbf{x}=[x_{0},...,x_{M-1}]\in 
\mathbf{R}^{M}$ is $K$-sparse in $\mathbf{R}^{M}$ if there exists a set of
indices $\{m_{1},...,m_{K}\}\subset \{0,...,M-1\}$, for small $K\ll M$ such
that: 
\begin{equation}
x_{m}=\left\{ 
\begin{array}{lll}
\neq 0 & if & m\in \{m_{1},...,m_{K}\} \\ 
=0 & if & m\notin \{m_{1},...,m_{K}\}
\end{array}
\right. .
\end{equation}
We consider the following encoding/decoding problem:

\begin{itemize}
\item  the sparse vector $\mathbf{x}\in \mathbf{R}^{M}$ is encoded in a
smaller dimensional vector $\mathbf{y}\in \mathbf{R}^{N}$ , $K<N\leq M$,
using a randomly generated $N\times M$ matrix $\mathbf{\Psi }$: 
\begin{equation}
\mathbf{y}=\mathbf{\Psi x}\in N\leq M,
\end{equation}
which is then submitted to a receiver;

\item  the receiver's decoding task consists in recovering the sparse vector 
$\mathbf{x}\in \mathbf{R}^{M}$, given the vector $\mathbf{y}\in \mathbf{R}%
^{N}$ and the random matrix $\mathbf{\Psi }$.
\end{itemize}

\noindent Thus, the coefficients of the vector $\mathbf{y}\in \mathbf{R}^{N}$
are the projections of the sparse vector $\mathbf{x}\in \mathbf{R}^{M}$ on
the vectors corresponding to the rows of the $N\times M$ matrix $\mathbf{%
\Psi }$. Reciprocally, we may say that $\mathbf{x}\in \mathbf{R}^{M}$
provides a sparse representation of $\mathbf{y}\in \mathbf{R}^{N}$ in the
redundant dictionary corresponding to the column vectors $\mathbf{\mathbf{%
\psi }}_{m}$ of the $N\times M$ random matrix $\mathbf{\Psi }$: 
\begin{equation}
\mathbf{\Psi =}[\mathbf{\mathbf{\psi }}_{0}|...|\mathbf{\mathbf{\psi }}%
_{M-1}]\mathbf{.}
\end{equation}
For convenience, we assume that the columns of the random matrix $%
\mathbf{\Psi }$ are normalized: 
\begin{equation}
\left\| \mathbf{\psi }_{m}\right\| _{2}=1,\quad m=0,...,M-1.
\end{equation}
Also, the analysis of algorithms for sparse signal recovery shows that the
matrix $\mathbf{\Psi }$ must satisfy the restricted isometry condition
[1-6]: 
\begin{equation}
(1-\delta _{x})\left\| \mathbf{x}\right\| _{2}^{2}\leq \left\| \mathbf{\Psi x%
}\right\| _{2}^{2}\leq (1+\delta _{x})\left\| \mathbf{x}\right\| _{2}^{2}.
\end{equation}
The restricted isometry constant $\delta _{x}$ is defined as the smallest
constant for which this property holds for all $K$-sparse vectors $\mathbf{x}%
\in \mathbf{R}^{M}$. In order to use this condition in practice, one would
need to be able to design a matrix satisfying the restricted isometry
condition. Recently it has been shown that one can generate such a matrix
with high probability, if the elements of the matrix are drawn independently
from certain probability distributions, such as a Gaussian distribution or a
Bernoulli distribution \cite{1}-\cite{6}. This is a consequence of the fact that in high
dimensions the probability mass of certain random variables concentrates
strongly around their expectation. Also, recent theoretic considerations
have shown that in order to achieve the restricted isometry condition, any $%
N\times M$ matrix $\mathbf{\Psi }$ must have at least $N\simeq cK\log (M/K)$
rows for some constant $c$ in order for the observation $\mathbf{y}=\mathbf{%
\Psi x}$ to allow an accurate reconstruction of $\mathbf{x}$ \cite{1}-\cite{6}.

Searching for the sparsest $\widetilde{\mathbf{x}}$ in the dictionary $%
\mathbf{\Psi }$ that matches $\mathbf{y}$ leads to the $l_{0}$ optimization
problem: 
\begin{equation}
\widetilde{\mathbf{x}}=\arg \min_{\mathbf{x}\in \mathbf{R}^{M}}\left\| 
\mathbf{x}\right\| _{0}\quad s.t.\quad \mathbf{\Psi x}=\mathbf{y}.
\end{equation}
Here, $\left\| \mathbf{x}\right\| _{0}$ is the $l_{0}$ norm, measuring the
number of nonzero coefficients in the vector $\mathbf{x}$. This
combinatorial optimization problem is NP-hard to solve and usually the
convexification of the objective function is introduced by replacing the $%
l_{0}$ norm with the $l_{1}$ norm [1-6] ($\left\| x\right\|
_{1}=\sum_{n=1}^{N}\left| x_{n}\right| $): 
\begin{equation}
\widetilde{\mathbf{x}}=\arg \min_{\mathbf{x}\in \mathbf{R}^{M}}\left\| 
\mathbf{x}\right\| _{1}\quad s.t.\quad \mathbf{\Psi x}=\mathbf{y}.
\end{equation}
The resulting optimization problem is known as Basis Pursuit (BP) and it can
be solved using linear programming techniques whose computational
complexities are polynomial \cite{1}-\cite{6}. However, the BP approach requires the
solution of a very large convex, nonquadratic optimization problem, and
therefore still suffers from high computational complexity. As an
alternative, here we consider an iterative greedy approach based on the MP
algorithm \cite{10}. MP tackles the problem by operating a local optimization,
as opposed to BP's global optimization strategy. MP has been proven to
achieve an accurate decomposition of the signal and it provides a
low-complexity alternative to BP, but requires an unbounded number of
iterations for convergence \cite{1}-\cite{7}, \cite{12}.

\section{The Matching Pursuit Algorithm}
\label{The Matching Pursuit Algorithm}

Matching Pursuit (MP) is an iterative heuristic algorithm which can be used
to obtain approximate solutions of the sparse recovery problem \cite{10}. Starting
from an initial approximation $\widetilde{\mathbf{x}}=0$ and residual $%
\mathbf{r}=\mathbf{y}$, the algorithm uses an iterative 'greedy' strategy to
pick the columns $\mathbf{\psi }_{m}$ of $\mathbf{\Psi }$ that are the most
strongly correlated with the residual. Then, successively their contribution
is subtracted from the residual, which this way can be made arbitrarily
small. The pseudo-code of the MP algorithm is:

\begin{enumerate}
\item  Initialize the variables: 
\begin{equation}
t\leftarrow 0,\widetilde{\mathbf{x}}\leftarrow 0,\mathbf{r}\leftarrow 
\mathbf{y},T.
\end{equation}

\item  While $\left\| \mathbf{r}\right\| _{2}>\varepsilon \left\| \mathbf{y}\right\| _{2}$ and $t<T$ repeat:

\begin{itemize}
\item  Find $i$ such that 
\begin{equation}
i\leftarrow \arg \max_{i\in \{1,...,M\}}|\left\langle \mathbf{r},\mathbf{\psi}_{i}\right\rangle| .
\end{equation}

\item  Update the estimate of the corresponding coefficient, the residual
and the iteration counter: 
\begin{equation}
\widetilde{x}_{i}\leftarrow \widetilde{x}_{i}+\left\langle \mathbf{r},%
\mathbf{\psi }_{i}\right\rangle ,
\end{equation}
\begin{equation}
\mathbf{r}\leftarrow \mathbf{r}-\left\langle \mathbf{r},\mathbf{\psi }%
_{i}\right\rangle \mathbf{\psi }_{i},
\end{equation}
\begin{equation}
t\leftarrow t+1.
\end{equation}
\end{itemize}

\item  Return $\widetilde{\mathbf{x}}.$
\end{enumerate}

\noindent The stopping criterion in the step 2 requires the residual to be
smaller than some fraction $\varepsilon $ of the 'target' $\mathbf{y}$.
Also, the computation stops if the number of iterations $t$ exceed the
maximum number allowed $T$. A shortcoming of the MP algorithm is that
although the asymptotic convergence is guaranteed and it can be easily
proven, the resulting approximation after any finite number of steps will in
general be suboptimal. This shortcoming can be corrected by using the
orthogonal version of MP, at a much higher computational cost \cite{5}, \cite{7}, \cite{12}.

For random dictionaries $\mathbf{\Psi }$, with dimensionality $N\times M$, each
inner product $\left\langle \mathbf{r},\mathbf{\psi }_{i}\right\rangle $
requires $N$ multiplications and $N-1$ additions. Each iteration requires $M$ such
inner product computations. Also, performing enough MP iterations to get a 
small reconstruction $\left\| \mathbf{r}\right\| _{2}<\varepsilon \left\| 
\mathbf{y}\right\| _{2}$ often means iterating $T\sim M$ times. Therefore,
the cost of computing the approximate solution $\widetilde{\mathbf{x}}$
could be as high as $O(N^{2}M^{2})>>O(N^{4})$, making MP very slow on
high-dimensional signals. Also, we should stress once again that in the case
of sparse signal recovery from random projections, the dictionaries are
completely random and therefore one cannot apply acceleration techniques
used for structured dictionaries (FFT, DCT, Gabor, Wavelet etc.) \cite{13}. Thus,
the hardware acceleration provided by a GPU implementation is probably the
only solution to speed up computation in this particular case.

In order to determine the effectiveness of the MP algorithm, we have conducted the following simple experiment. 
The sparse signals $\mathbf{x}\in \mathbf{R}^{M}$ were generated by drawing the $K$
non-zero components from a uniform distribution on $[-1,+1]$. Also, the
random $N\times M$ matrix $\mathbf{\Psi }$ was drawn from a Bernoulli
distribution: 
\begin{equation}
\psi _{nm}=\left\{ 
\begin{array}{ll}
+1/\sqrt{N} & with\ probability\ 1/2 \\ 
-1/\sqrt{N} & with\ probability\ 1/2
\end{array}
\right. .
\end{equation}
We have considered two parameters $\sigma =K/M$ and $\rho =N/M$ and we have
computed the relative reconstruction error: 
\begin{equation}
E(\sigma ,\rho )=\left\| \widetilde{\mathbf{x}}-\mathbf{x}\right\|
_{2}^{2}\left\| \mathbf{x}\right\| _{2}^{-2}\times 100\%,
\end{equation}
as a function of $\sigma $ and $\rho $.

\begin{figure}
\centerline{
\mbox{\includegraphics[width=3.00in]{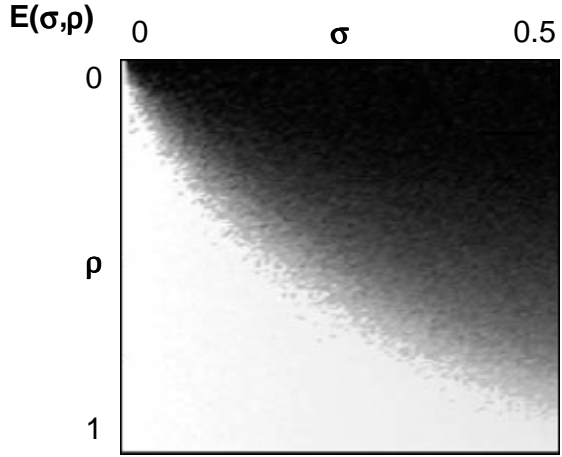}} }
\caption{Sparse reconstruction error of the MP algorithm 
(the error grows from white ($0\%$) to black ($100\%$)).}
\label{Fig1}
\end{figure}

The results for $M=1000$, averaged over $100$ trials, are shown in Figure 1.
The stopping parameters were set to $\varepsilon =10^{-7}$ and respectively $%
T=N$. The numerical results are in very good agreement with the previous
theoretical considerations, showing a logarithmic dependence between $\rho $
and $\sigma $. For a small number of projections (small $\rho $) the
algorithm is unstable and the reconstruction error grows. By increasing the
number of projections (large $\rho $) the MP algorithm is able to recover
exactly the $K$-sparse signal $\mathbf{x}\in \mathbf{R}^{M}$.

\section{CPU vs GPU (BLAS vs CUBLAS)}
\label{CPU vs GPU (BLAS vs CUBLAS)}

Due to their tremendous memory bandwidth and computational horsepower, GPUs
are becoming an efficient alternative to solve computer-intensive
applications. The newly developed GPUs now include fully programmable
processing units that follow a stream programming model and support
vectorized single and double precision floating-point operations. For
example, the CUDA computing environment provides a standard C like language
interface to the NVIDIA GPUs \cite{8}. The computation is distributed into
sequential grids, which are organized as a set of thread blocks. The thread
blocks are batches of threads that execute together, sharing local memories
and synchronizing at specified barriers. An enormous number of blocks, each
containing maximum 512 threads, can be launched in parallel in the grid.

In this paper we make use of CUBLAS, a recent implementation of BLAS,
developed by NVIDIA on top of the CUDA programming environment \cite{9}. CUBLAS
library provides functions for:
\begin{itemize}
\item  creating and destroying matrix and vector objects in GPU memory;

\item  transferring data from CPU mainmemory to GPU memory;

\item  executing BLAS on the GPU;

\item  transferring data from GPU memory back to the CPU mainmemory.
\end{itemize}

BLAS defines a set of fundamental operations on vectors and matrices which
can be used to create optimized higher-level linear algebra functionality:

\begin{itemize}
\item  Level 1 BLAS perform scalar, vector and vector-vector operations;

\item  Level 2 BLAS perform matrix-vector operations;

\item  Level 3 BLAS perform matrix-matrix operations.
\end{itemize}
\noindent Highly efficient implementations of BLAS exist for most current
computer architectures and the specification of BLAS is widely adopted in
the development of high quality linear algebra software, such as LAPACK,
IMKL and GNU Scientific Library (GSL) \cite{11}. We selected GSL CBLAS, for our
host (CPU) implementation, due to its portability on various platforms
(Windows/Linux/OSX, Intel/AMD) and because it is free and easy to use in
combination with GCC (GNU Compiler). The GSL library provides a low-level
layer which corresponds directly to the C-language BLAS standard, referred
here as CBLAS, and a higher-level interface for operations on GSL vectors
and matrices \cite{11}.

The CBLAS and CUBLAS implementations of the MP algorithm require the
following functions/kernels:
\bigskip

\noindent CBLAS: gsl\_blas\_sgemv, gsl\_blas\_dgemv

\noindent CUBLAS: cublasSgemv, cublasDgemv

\begin{itemize}
\item  compute in single/double precision the matrix-vector product and
sum: 
\begin{equation}
\mathbf{y=}\alpha \mathbf{Ax}+\beta \quad or\quad \mathbf{y=}\alpha \mathbf{A%
}^{T}\mathbf{x}+\beta \mathbf{y.}
\end{equation}
\end{itemize}

\noindent \noindent CBLAS: gsl\_blas\_isamax, gsl\_blas\_idamax

\noindent CUBLAS: cublasIsamax, cublasIdamax

\begin{itemize}
\item  return the smallest index of the maximum magnitude element of the
single/double precision vector x: 
\begin{equation}
m=\arg \max_{m}\left| x_{m}\right| .
\end{equation}
\end{itemize}

\noindent \noindent CBLAS: int gsl\_blas\_saxpy, gsl\_blas\_daxpy

\noindent CUBLAS: cublasSaxpy, cublasDaxpy

\begin{itemize}
\item  compute the single/double precision sum 
\begin{equation}
\mathbf{y}=\alpha \mathbf{x}+\mathbf{y}.
\end{equation}
\end{itemize}

\noindent CBLAS: gsl\_blas\_snrm2, gsl\_blas\_dnrm2

\noindent CUBLAS: cublasSnrm2, cublasDnrm2

\begin{itemize}
\item  compute the Euclidean norm of a single/double precision vector: 
\begin{equation}
\left\| \mathbf{x}\right\| _{2}=\sqrt{\sum_{m=0}^{M}x_{m}^{2}}.
\end{equation}
\noindent 
\end{itemize}

\noindent These are the critical functions/kernels, which are efficiently
exploited in the parallel CUBLAS implementation. The other involved
functions are for vector/matrix memory allocation and vector/matrix
accessing, device (GPU) initialization, host-device data transfer and error
handling. We stress once again that in the CUBLAS implementation the data
space is allocated both on host (CPU) mainmemory and on device (GPU) memory.
After the data is initialized on host it is transferred on device, where the
main parallel computation occurs. The results are then transferred back on
host memory where the solution can be checked against the reference. The
data transfer from host to device and back is not critical for MP. For
example the host to device dictionary transfer only takes place once, since
all signals are encoded/decoded with the same dictionary. The time for
signal transfer from host to device and back is insignificant comparing to
the device computation time. For convenience, we have included also a timer
which measures the time of all main operations involved by the MP algorithm.
The double precision code for CBLAS and CUBLAS implementations is given in
Appendix 1 and respectively Appendix 2. These implementations can be easily
modified in order to meet end user's needs. For example the single precision
BLAS data allocation and vector/matrix accessing functions have the prefix
gsl\_vector\_float, gsl\_matrix\_float etc. (see \cite{9}, \cite{11} for details).

The numerical experiments have been carried out on the following system
custom build by the author: CPU: AMD Phenom 9950 (2.6GHz); Motherboard: ASUS
M3N78 Pro (chipset GeForce m8300); RAM 8Gb DDR2 800MHz; On-board GPU:
8400GS, 512 Mb DDR2 (shared); PCI-E GPU: XFX GTX280, 1024 Mb DDR3; NVIDIA
Linux 64-bit driver (177.67); CUDA Toolkit and SDK 2.0; Ubuntu Linux 64-bit
8.04.1, GNU Scientific Library v.1.11; Compilers: GCC (GNU), NVCC (NVIDIA).
\noindent This system configuration gives the opportunity to evaluate the
performance of the GPU code on two different GPUs. The first GPU (GPU0) is
the core (G86) of the GeForce 8400GS, which is used as an entry level,
on-board solution for the GeForce m8300 motherboard, it has 16 stream
processors and 512 Mb DDR2 RAM (shared), and supports only single precision
floating-point operations. The second GPU (GPU1) is the core (GT200) of
GeForce GTX280, a high end solution with 240 stream processors and 1Gb DDR3
RAM, which supports both single and double precision and it is theoretically
capable of 1 Tflop computational power.

In order to generate the same random dictionaries and signals, the seed of
the random number generator (srand()) was set to the same value in both
CBLAS and CUBLAS implementations. We varied the length $M$ of the signal and
the size of the dictionary from $M=1000$ to $M=15000$. The number of
non-zero elements in the sparse signal was fixed to $K=\left\lfloor
0.07M\right\rfloor $ ($\sigma =0.07$) and the number of projections was set
to $N=\left\lfloor M/2\right\rfloor $ ($\rho =0.5$). Also, the maximum
number of iterations and the reconstruction error were set to $T=N$ and
respectively $\varepsilon =10^{-7}$. With these parameters, the MP algorithm
is able to find the solution with an error $\varepsilon =10^{-7}$ in maximum 
$N$ iteration steps.

\begin{figure}
\centerline{
\mbox{\includegraphics[width=3.00in]{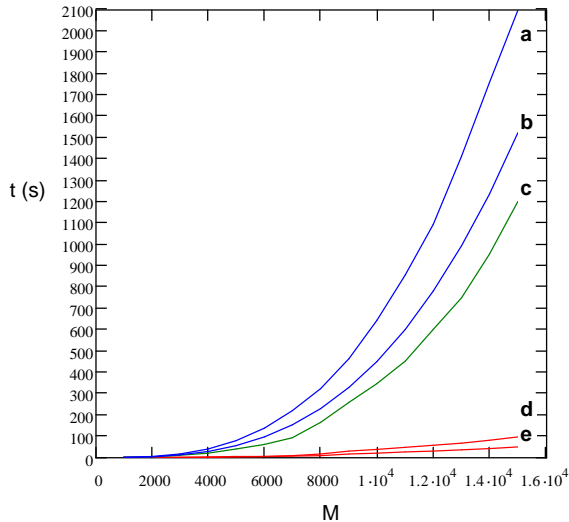}} }
\caption{Numerical results for CPU and GPU computation: 
a - CPU double precision; b - CPU single precision; 
c - GPU0 single precision; d - GPU1 double precision; e - GPU1 single precision.}
\label{Fig2}
\end{figure}

In Figure 2 we give the results obtained for single and respectively double
precision. It is interesting to note that for small dictionaries $M\leq 2000$ the CPU is
actually faster than the GPU0. The gap between CPU and GPU1 increases very
fast by increasing the size of the dictionary. The GPU1 versus CPU speed
reaches a maximum of 31 times in single precision and respectively 21 times
in double precision, for $M=15,000$. These results shows once again that GPU
performance is dependent on the scale of the problem. Thus, in order to
exploit efficiently the massive parallelism of GPUs and to effectively use
the hardware capabilities, the problem itself needs to scale accordingly,
such that thousands of threads are defined and used in computation.

The current version of CUBLAS is not yet perfectly tuned
comparing to the highly optimized CBLAS. This is revealed by using a
logarithmic time scale representation of the previous results. In Figure 3 one
can see that the CPU computation time is monotonously and smoothly
increasing with the size of the problem (as expected) for the CBLAS code,
while the GPU running time exhibits a discontinuity around $M=8192$, where
the performance drops quite sharply. This discontinuity is present for both
GPUs and also for both single and double precision implementations. Thus, we
conclude that it is not a hardware artifact but rather a CUBLAS
implementation artifact.

\begin{figure}
\centerline{
\mbox{\includegraphics[width=3.05in]{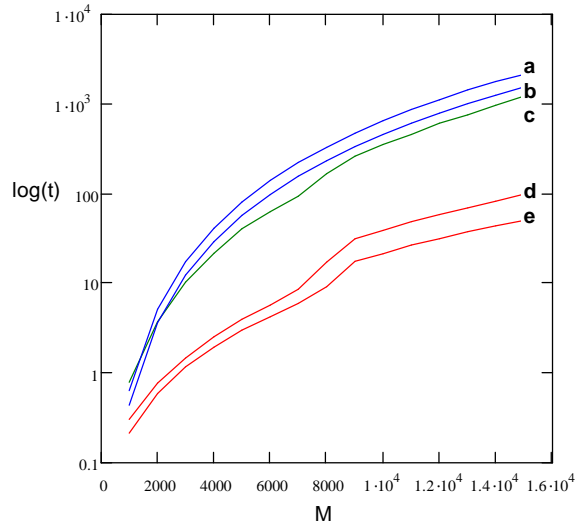}} }
\caption{Numerical results for CPU and GPU computation in logarithmic time scale: 
a - CPU double precision; b - CPU single precision; c - GPU0 single precision; 
d - GPU1 double precision; e - GPU1 single precision.}
\label{Fig3}
\end{figure}

\section{Conclusion}
\label{Conclusion}

In this paper we have presented a comparison between the CBLAS (CPU) and the
CUBLAS (GPU) implementations of the MP algorithm, a frequently used method
to recover sparse signals from random projections. The MP algorithm
efficiently exploits the high computing power of the GPU. For large
problems, the GPU code is up to 31 (21) times faster, in single (double)
precision, than the CPU code. For example, the sparse reconstruction time
for a signal of a length $M=16,384$ encoded with $N=8,192$ random vectors
takes in average $1,900$ s on a CPU, while the same decoding problem takes
less than $58$ s on the GPU. This kind of performance improvement makes it
possible to approach even more computational hungry algorithms like the
orthogonal version of MP, which is our next target for GPU
implementation.


\footnotesize{

\section*{Appendix 1}
\begin{verbatim}
/*
 *  Double precision CBLAS (GSL) implementation  
 *  of Matching Pursuit algorithm for 
 *  sparse signal recovery from random projections
*/
 
/* Includes, system */
#include <stdio.h>
#include <math.h>
#include <time.h>
 
/* Includes, GSL & CBLAS */
#include <gsl/gsl_vector.h>
#include <gsl/gsl_matrix.h>
#include <gsl/gsl_blas.h>
 
 /* Number of columns and rows in dictionary */
#define M (4000)
#define N ((int)(M/2))
 
/* Number of non-zero elements in sparse signal */
int K = 0.07*M;
 
/* Residual error */
double epsilon = 1.0e-7;
 
/* Maximum number of iterations */
int T = N;
 
/* Sign function */
double sign(double x){return (x>=0) - (x<0);}
 
int main(int argc, char** argv)
 {
 int n, m, k, t, q;
 double normi, normf, a , norm = sqrt(N), htime;
 
 printf(''\nProblem dimensions: 
        NxM=%dx%d, K=%d'', N, M, K);
 
 /* Initialize srand and clock */
 srand(time(NULL));
 clock_t start = clock();
 
 /* Initiallize Bernoulli random dictionary */
 gsl_matrix *D = gsl_matrix_alloc (N, M);
 for(m=0; m<M; m++) 
  {
  for(n=0; n<N; n++) 
   {
   a = sign(2.0*rand()/(double)RAND_MAX-1.0)/norm;
   gsl_matrix_set (D, n, m, a);
   }
  }
 
 /* Create a random K-sparse signal */
 gsl_vector *x = gsl_vector_alloc(M);
 for(k=0; k<K; k++)
  {
  gsl_vector_set(x, rand()%M, 2.0*rand()
                 /(float)RAND_MAX - 1.0);
  }
 
 /* Allocate memory for solution */
 gsl_vector *z = gsl_vector_calloc(M);
 
 /* Allocate memory for residual */
 gsl_vector *r = gsl_vector_calloc(M);
 
 /* Allocate memory for the encoding vector */
 gsl_vector *y = gsl_vector_alloc(N);
 htime = ((double)clock()-start)/CLOCKS_PER_SEC;
 printf(''\nTime for data allocation: %f'', htime);
 
 /* Encoding the signal */ 
 start = clock();   
 gsl_blas_dgemv(CblasNoTrans, 1.0, D, x, 0.0, y);
 htime = ((double)clock()-start)/CLOCKS_PER_SEC;
 printf(''\nTime for encoding: %f'', htime); 
 
 /* Decoding the signal */
 start = clock();   
 normi = gsl_blas_dnrm2(y);
 epsilon = sqrt(epsilon * normi);
 normf = normi;
 t = 0;
 while(normf > epsilon && t < T)
  {
  gsl_blas_dgemv(CblasTrans, 1.0, D, y, 0.0, r);
  q = gsl_blas_idamax(r);
  a = gsl_vector_get(r, q);
  gsl_vector_set(z, q, gsl_vector_get(z, q) + a);
  gsl_blas_daxpy(-a, 
                 &gsl_matrix_column(D, q).vector, 
                 y);
  normf = gsl_blas_dnrm2(y);
  t++;
  }
 htime = ((double)clock()-start)/CLOCKS_PER_SEC;
 printf(''\nTime for decoding: %f'', htime); 
 a = 100.0*(normf*normf)/(normi*normi);
 printf(''\nComputation residual error: %f '', a);
 
 /* Check the solution against the reference*/
 printf(''\nSolution (first column), 
        Reference (second column):'');
 getchar();
 for(m=0; m<M; m++)
  {
  printf(''\n%f\t%f'',  gsl_vector_get(x, m), 
         gsl_vector_get(z, m));
  }
 normi = gsl_blas_dnrm2(x);
 gsl_blas_daxpy(-1.0, x, z);
 normf = gsl_blas_dnrm2(z);
 a = 100.0*(normf*normf)/(normi*normi);
 printf(''\nSolution residual error: %f \n'', a);
 
 /* Memory clean up and shutdown*/
 gsl_vector_free(y);
 gsl_vector_free(r);
 gsl_vector_free(z);
 gsl_vector_free(x);
 gsl_matrix_free(D);
 return EXIT_SUCCESS;
 }
\end{verbatim}

\section*{Appendix 2}
\begin{verbatim}
/*
 *  Double precision CUBLAS (NVIDIA) implementation 
 *  of Matching Pursuit algorithm for 
 *  sparse signal recovery from random projections.
*/
 
/* Includes, system */
#include <stdio.h>
#include <stdlib.h>
#include <string.h>
#include <time.h>
 
/* Includes, cuda */
#include <cublas.h>
 
/* Number of columns and rows in dictionary */
#define M (2000)
#define N ((int)(M/2))
 
/* Number of non-zero elements in sparse signal */
int K = 0.07*M;
 
/* Residual error */
double epsilon = 1.0e-7;
 
/* Maximum number of iterations */
int T = N;
 
/* Sign function */
double sign(double x){return (x>=0) - (x<0);}
 
/* Matrix indexing convention */
#define id(m, n, ld) (((n) * (ld) + (m)))
 
int main(int argc, char** argv)
 { 
 cublasStatus status;
 double *h_D, *h_X, *h_C, *c; 
 double *d_D = 0, *d_S = 0, *d_R = 0; 
 int MN = M*N, m, n, k, q, t;
 double norm=sqrt(N), normi, normf, a, dtime;
 
 printf(''\nProblem dimensions: 
        NxM=%dx%d, K=%d'', N, M, K);
 
 /* Initialize srand and clock */
 srand(time(NULL));
 clock_t start = clock(); 
 
 /* Initialize cublas */
 status = cublasInit();
 if (status != CUBLAS_STATUS_SUCCESS) 
  {
  fprintf(stderr, 
          ''! CUBLAS initialization error\n'');  
  return EXIT_FAILURE;
  }
 
 /* Initialize dictionary on host */
 h_D = (double*)malloc(MN * sizeof(h_D[0]));
 if(h_D == 0) 
  {
  fprintf(stderr, ''! host memory allocation error 
         (dictionary)\n'');  
  return EXIT_FAILURE;
  }
 for(n = 0; n < N; n++) 
  {
  for(m = 0; m < M; m++) 
   {
   a = sign(2.0*rand()/(double)RAND_MAX-1.0)/norm;
   h_D[id(m, n, M)] = a; 
   }
  }
 
 /* Create a random K-sparse signal */ 
 h_X = (double*)calloc(M, sizeof(h_X[0]));
 if(h_X == 0) 
  {
  fprintf(stderr, ''! host memory allocation 
          error (signal)\n''); 
  return EXIT_FAILURE;
  }
 for(k=0; k<K; k++) 
  {
  a = 2.0*rand()/(double)RAND_MAX - 1.0;
  h_X[rand()%M] = a; 
  }
 
 /* Allocate solution memory on host */
 h_C = (double*)calloc(M, sizeof(h_C[0]));
 if(h_C == 0) 
  {
  fprintf(stderr, ''! host memory allocation 
          error (solution)\n''); 
  return EXIT_FAILURE;
  }
 c = (double*)calloc(1, sizeof(c));
 if(c == 0) 
  {
  fprintf(stderr, ''! host memory allocation 
          error (c)\n''); 
  return EXIT_FAILURE;
  }
 dtime = ((double)clock()-start)/CLOCKS_PER_SEC;
 printf(''\nTime for host data allocation: 
        %f'', dtime);  
 start=clock();    
 
 /* Host to device data transfer: dictionary */
 status = cublasAlloc(MN, sizeof(d_D[0]), 
          (void**)&d_D);
 if(status != CUBLAS_STATUS_SUCCESS) 
  {
  fprintf(stderr, ''! device memory allocation 
          error (dictionary)\n''); 
  return EXIT_FAILURE;
  }    
 status = cublasSetVector(MN, sizeof(h_D[0]), 
                          h_D, 1, d_D, 1);
 if(status != CUBLAS_STATUS_SUCCESS) 
  {
  fprintf(stderr, ''! device access error 
          (write dictionary)\n''); 
  return EXIT_FAILURE;
  }
 
 /* Host to device data transfer: signal */
 status = cublasAlloc(M, sizeof(d_R[0]), 
                      (void**)&d_R);
 if(status != CUBLAS_STATUS_SUCCESS) 
  {
  fprintf(stderr, ''! device memory allocation 
          error (signal)\n''); 
  return EXIT_FAILURE;
  }     
 status = cublasSetVector(M, sizeof(h_X[0]), 
                          h_X, 1, d_R, 1);
 if(status != CUBLAS_STATUS_SUCCESS) 
  {
  fprintf(stderr, ''! device access error 
          (write signal)\n''); 
  return EXIT_FAILURE;
  }
 status = cublasAlloc(N, sizeof(d_S[0]), 
                      (void**)&d_S);
 if(status != CUBLAS_STATUS_SUCCESS) 
  {
  fprintf(stderr, ''! device memory allocation 
          error (projected vector)\n''); 
  return EXIT_FAILURE;
  }     
 dtime = ((double)clock()-start)/CLOCKS_PER_SEC;
 printf(''\nTime for Host to Device data transfer: 
        %f (s)'', dtime);  
 
 /* Encoding the signal on device*/      
 start = clock();      
 cublasDgemv('t', M, N, 1.0, d_D, M, 
             d_R, 1, 0.0, d_S, 1);
 status = cublasGetError();
 if(status != CUBLAS_STATUS_SUCCESS) 
  {
  fprintf(stderr, ''! kernel execution 
          error (encoding)\n''); 
  return EXIT_FAILURE;
  }
 dtime = ((double)clock()-start)/CLOCKS_PER_SEC;
 printf(''\nTime for encoding: %f (s)'', dtime);  
 
 /* Decoding the signal on device*/      
 start = clock();      
 normi = cublasDnrm2 (N, d_S, 1);  
    epsilon = sqrt(epsilon*normi); 
    normf = normi;
    t = 0; 
 while(normf > epsilon && t < T)
  {
  cublasDgemv('n', M, N, 1.0, d_D, M, 
              d_S, 1, 0.0, d_R, 1);
  q = cublasIdamax (M, d_R, 1) - 1;
  cublasGetVector(1, sizeof(c), 
                  &d_R[q], 1, c, 1); 
  h_C[q] = *c + h_C[q];
  cublasDaxpy (N, -(*c), &d_D[q], M, d_S, 1);
  normf = cublasDnrm2 (N, d_S, 1);  
  t++;
  }
 status = cublasGetError();
 if(status != CUBLAS_STATUS_SUCCESS) 
  {
  fprintf(stderr, ''! kernel execution 
          error (decoding)\n''); 
  return EXIT_FAILURE;
  }
 dtime = ((double)clock()-start)/CLOCKS_PER_SEC;
 printf(''\nTime for decoding: %f (s)'', dtime);  
 a = 100.0*(normf*normf)/(normi*normi);
 printf(''\nComputation residual error: %f'', a);
 
 /* Check the solution */
 printf(''\nSolution (first column), 
        Reference (second column):''); 
 getchar();
 for(m=0; m<M; m++) 
  {
  printf(''\n%f\t%f'',  h_C[m], h_X[m]);
  }
 normi = 0; normf = 0; 
 for(m=0; m<M; m++)
  {
  normi = normi + h_X[m]*h_X[m]; 
  normf = normf + 
          (h_C[m] - h_X[m])*(h_C[m] - h_X[m]);
  } 
 printf(''\nSolution residual error: 
        %f'', 100.0*normf/normi); 
 
 /* Memory clean up */
 free(h_D);
 free(h_X);
 free(h_C);
 status = cublasFree(d_D);
 status = cublasFree(d_S);
 status = cublasFree(d_R);
 if(status != CUBLAS_STATUS_SUCCESS) 
  {
  fprintf(stderr, 
          ''! device memory free error\n''); 
  return EXIT_FAILURE;
  }
 
 /* Shutdown */
 status = cublasShutdown();
 if(status != CUBLAS_STATUS_SUCCESS)  
  {
  fprintf(stderr, 
          ''! cublas shutdown error\n''); 
  return EXIT_FAILURE;
  }
 if(argc<=1 || strcmp(argv[1],''-noprompt'')) 
  {
  printf(''\nPress ENTER to exit...\n'');  
  getchar();
  }
 return EXIT_SUCCESS;
}
\end{verbatim}
}

\begin{thebibliography}{99}

\bibitem{1}  Cand\`{e}s, E., Tao, T., ``Near Optimal Signal Recovery from Random
Projections: Universal Encoding Strategies?'', {\it IEEE Trans. Inf. Theory}, V52, pp. 5406-5425, 2006.

\bibitem{2}  Haupt, J., Nowak, R., ``Signal Reconstruction from Noisy Random Projections'', 
{\it IEEE Trans. Inf. Theory}, V52, pp. 4036-4048, 2006.

\bibitem{3}  Donoho, D., ``Compressed Sensing'', {\it IEEE Trans. Inf. Theory}, V52, pp. 1289-1306, 2006.

\bibitem{4}  Cand\`{e}s, E., ``Compressive sampling,'' 
{\it Int. Congress of Mathematics}, Madrid, Spain, V3, pp. 1433-1453, 2006.

\bibitem{5}  Tropp, J., Gilbert, A., ``Signal Recovery from Random Measurements via Orthogonal Matching Pursuit'', 
{\it IEEE Trans. Inf. Theory}, V53, pp. 4655-4666 2007.

\bibitem{6}  Baraniuk, R., ``Compressive sensing'', {\it IEEE Signal Proc. Mag.}, 24, pp. 118-121, 2007.

\bibitem{7}  Andrecut, M., Kauffman, S. A., ``On the Sparse Reconstruction of Gene Networks'', 
{\it J. Comput. Biol.}, V15, pp. 21-30, 2008.

\bibitem{8}  NVIDIA, {\it Nvidia CUDA Compute Unified Device Architecture},
Programming Guide, 2008.

\bibitem{9}  NVIDIA, {\it CUBLAS Library}, 2008.

\bibitem{10}  Mallat, S., Zhang, Z., ``Matching Pursuit in a Time-Frequency Dictionary'', 
IEEE Trans. Signal Process., V41, pp. 3397-3415, 1993.

\bibitem{11}  Galassi M. et al., 
{\it GNU Scientific Library Reference Manual - Revised Second Edition}, Network Theory Limited, 2006.

\bibitem{12}  Pati, Y. C., Rezaiifar, R., Krishnaprasad, P. S., 
``Orthogonal Matching Pursuit: Recursive Function Approximation with Applications to Wavelet Decomposition'', 
{\it Proc. 27th Asilomar Conf. Signals Syst. Comput.}, V1, pp. 40-44, 1993.

\bibitem{13}  Krstulovic, S., Gribonval, R., ``MPTK: Matching Pursuit made Tractable'', 
{\it Proc. Int. Conf. Acoust. Speech Signal Process. (ICASSP)}, V3, pp. III496-III499, 2006.

\end{thebibliography}
\end{document}